\documentclass[preprint,aps,showpacs,prd,tightenlines,twocolumn,10pt]{revtex4}

\usepackage[latin1]{inputenc}
\usepackage{bbm}

\newcommand{\be}{\begin{eqnarray}} % only untightened
\newcommand{\ee}{\end{eqnarray}}
\newcommand{\nn}{~\nonumber \\}

\newcommand{\ssh}{\hskip 0.6mm\not\hskip -0.6mm}
\newcommand{\SSH}{\hskip 0.9mm\not\hskip -0.9mm}

\newcommand{\bmp}{\noindent\begin{minipage}{16cm}}
\newcommand{\emp}{\end{minipage}\vskip 7mm} % 7mm untightened

\begin{document}

%%%%%%%%%%%%%%%%%%%%%%%%%%%%%%%%%%%%%%%%%%%%%%%%%%%%%%%%%%%%%

\title{Lorentz invariant ensembles of vector backgrounds}

\author{Dennis D. Dietrich}
\affiliation{The Niels Bohr Institute, Copenhagen, Denmark}
\author{Stefan Hofmann}
\affiliation{Perimeter Institute for Theoretical Physics, Waterloo,
Ontario, Canada}

\date{June 22, 2005}

%%%%%%%%%%%%%%%%%%%%%%%%%%%%%%%%%%%%%%%%%%%%%%%%%%%%%%%%%%%%%

\begin{abstract}

We consider gauge field theories in the presence of ensembles of vector
backgrounds. While Lorentz invariance is explicitely broken in the 
presence of any 
single background, here, the Lorentz invariance of the theory is restored by
averaging over a Lorentz invariant ensemble of backgrounds, i.e. a set of
background vectors that is mapped onto itself under Lorentz transformations. 
This framewkork is used to study the effects of a non-trivial but Lorentz 
invariant vacuum structure or mass dimension two vector condensates by
identifying the background with a shift of the gauge field. Up to now, the 
ensembles used in the literature comprise configurations corresponding to
non-zero field tensors together with such with vanishing field strength.  
We find that even when constraining the ensembles to pure gauge
configurations, the usual high-energy degrees of freedom are removed from the 
spectrum of asymptotic states in the presence of said backgrounds in
euclidean and in Minkowski space. We establish this result not only for 
the propagators to all orders in the background and otherwise at tree level
but for the full propagator.

~\\

\noindent
Keywords: Lorentz invariance, classical and semiclassical methods in gauge field theories

\pacs{
11.30.Cp, % Lorentz & Poincare invariance
11.15.Kc, % Classical and semiclassical methods in gauge field theories
03.30.+p, % Special relativity
12.60.-i  % beyond SM
}

\end{abstract}

%%%%%%%%%%%%%%%%%%%%%%%%%%%%%%%%%%%%%%%%%%%%%%%%%%%%%%%%%%%%%%

\maketitle

%%%%%%%%%%%%%%%%%%%%%%%%%%%%%%%%%%%%%%%%%%%%%%%%%%%%%%%%%%%%%

\section{Introduction}

Vector backgrounds appear in numerous sectors of physics. For example they
can be used to include the influence of mass dimension two condensates into
quantum chromodynamics (QCD) by shifting the gauge field and subsequently
restoring Lorentz invariance by averaging over a Lorentz invariant ensemble
of backgrounds \cite{hp,ls}. There this construction removes quarks and gluons 
from the spectrum of freely propagating particles. Lately, those condensates 
have attracted much attention in various respects (see e.g. \cite{a2}).

Analogous vector terms also occur in other domains. Introduced without the 
subsequent restoration of Lorentz invariance they are used in extensions of 
the standard model
explicitely breaking Lorentz invariance \cite{lv}. The inclusion of such
terms is motivated from string theory \cite{string} and non-commutative field 
theory \cite{noncom}. Lorentz invariance is one of the best-tested 
postulates of field theory but the current experimental status still leaves 
room for small deviations. However the approach under
investigation in this article is Lorentz invariant. 

We also identify the background vector with a shift of the gauge field. First 
however, in section II, we discuss the general framework for the 
inclusion of a dependence on an additional vector into a previously Lorentz 
invariant theory which {\it a priori} breaks Lorentz invariance explicitely 
but where the correlators are defined as the ensemble average over a Lorentz 
invariant set. Thereby a modified theory is obtained which preserves Lorentz 
invariance. A Lorentz invariant ensemble is a set of vectors which is mapped 
onto itself under any Lorentz transformation, while, of course, almost every 
single element changes.
As a next step, in section IIA, we carry out a classification of 
the weight functions which characterise those ensembles. At variance with
euclidean space some subtleties arise 
in Minkowski space. Commonly the used sets contain configurations
leading to vanishing and non-vanishing field tensors \cite{hp,ls}. Here, we 
limit the ensembles further by
constraining them to pure gauge configurations. In this framework we
analyse the objects central to the modified theory, i.e., the generating
functional for the Green functions in section IIB and the fermionic
two-point function by solving its equation of motion explicitly \cite{ddd} in 
section IIC. 

Finally, in section III, we summarise the
paper, the main result being that the propagation of fermions over
arbitrarily long distances is already stopped in ensembles of pure gauge
configurations of the background for euclidean and Minkowski spaces
characterised by their respective metrics.
This result is not only derived to all orders in the background and
otherwise at tree level but for the exact propagator.

Other observations are non-gaussianity of the resulting theory, structural 
similarities between the present approach 
and Lorentz invariant generalisations of chemical potentials as well as a
technical relationship to stochastic field theory. Last but not least, in
Minkowski space the modification of the fermionic two-point Green function
amounts to a contribution of a scalar to the fermion's self energy but
without external legs. This again indicates in a diagrammatic way that the 
ultraviolet degrees of freedom are removed from the asymptotic spectrum.

%%%%%%%%%%%%%%%%%%%%%%%%%%%%%%%%%%%%%%%%%%%%%%%%%%%%%%%%%%%%

\section{Breaking and restoring Lorentz invariance}

Regard a gauge field theory which is modified by including a dependence on a
vector $\Phi$. The translational invariance of the system remains
intact, because the vector is constant. The Lorentz invariance of the theory 
is to be restored by
taking the average over an ensemble of vectors $\Phi$ characterised by a
Lorentz invariant weight \mbox{$W(\Phi)$} \footnote{Integrations over the
$\mathbbm{R}^4$ are denoted by a subscript, e.g.: 
\mbox{$\int d^4\Phi=:\int_\Phi$}.}:
\be
\left<{\cal O}\right>_W
=
\int_\Phi W(\Phi)~{\cal O},
\ee
where ${\cal O}$ stands for a generic operator, here and in the following,
and with the normalisation condition:
\be
\int_\Phi W(\Phi)=1.
\label{normalisation}
\ee
Apart from the case where \mbox{$\Phi=0$}, which corresponds to the original
theory, functions of $\Phi^2$ are the only Lorentz invariant
quantities that can be constructed from the vector $\Phi$. The most general 
Lorentz invariant weight \mbox{$W(\Phi)$} is given 
by the sum of an arbitrary normalisable function \mbox{$w=w(\Phi^2)$} and a 
delta distribution $\delta^{(4)}(\Phi)$:
\be
W(\Phi)=c\delta^{(4)}(\Phi)+w(\Phi^2)
\label{weight}
\ee
In \cite{hp} the vector $\Phi$ represent a vector condensate translating the
gauge boson field \mbox{$A\rightarrow A+\Phi$}. That system is investigated 
with a
euclidean metric for quantum chromodynamics (QCD). In the sense \mbox{$c=0$} 
the
weight chosen there (\mbox{$\sim\exp\{-\Phi^2/\Lambda^2\}$}) does not contain 
the
unmodified theory. This manifests itself in the one-particle pole being
removed from the quark and gluon propagators determined to all orders in
$\Phi$, meaning that there the partons do not propagate over arbitrarily 
long distances. They are no longer
part of the asymptotic spectrum. The lowest order in an expansion for
momenta large compared to the scale $\Lambda^2$ reproduces the standard free
propagators. 

In QCD the vector $\Phi$ also carries colour indices:
\mbox{$\Phi^2=\Phi^a_\mu\Phi^{a\mu}$}. 
The vector $\Phi$ is to transform
homogeneously under gauge transformations whence any function of $\Phi^2$ is
gauge covariant. 
This also establishes the connection of the present approach with mass 
dimension two condensates because now $\Phi$ acts as a
contribution to the gauge field \cite{hp,ls,a2}.
Due to the non-abelian nature of the gauge theory, the
ensemble of constant vectors characterised by a function of its square
contains members which are pure gauge configurations and such leading to a
non-vanishing field tensor. In our investigation, we will distinguish these
cases by limiting the ensemble to vanishing field tensors. This in itself is
a gauge invariant criterion. For the fermionic sector this leads to an
analogue of quantum electrodynamics (QED) which we will study in the following.
Further, we will compare the results for the different metrics.

% % % % % % % % % % % % % % % % % % % % % % % % % % % % % % % % % % % % % %

\subsection{Weight classification}

Let us begin with a classification of the weight functions. In principle, in 
euclidean space the case \mbox{$\Phi=0$} is already included in $w(\Phi^2)$ 
as there \mbox{$\Phi^2=0$} implies \mbox{$\Phi=0$}. Nevertheless, in 
order to mark the potential contribution from the unmodified theory clearly, 
i.e., from \mbox{$\Phi=0$}, let us split it off in form of a delta 
distribution in accordance with Eq. (\ref{weight}). The normalisation 
condition (\ref{normalisation}) then implies
\be
\pi^2\int_0^{+\infty} v~dv~w_{\mathrm E}(v)=1-c
\label{normalisationeuclid}
\ee
with \mbox{$v:=\Phi^2$} and where the subscript $_E$ marks the Euclidean
case.

Every possible Lorentz invariant weight function 
\mbox{$w_{\mathrm E}(\Phi^2)$} can be reconstructed by a convolution with a 
delta weight
\be
w_{\mathrm E}(\Phi^2)
=
\int d\lambda~\delta(\Phi^2-\lambda)~w_{\mathrm E}(\lambda).
\ee
In this sense the delta weight:
\be
w_{\mathrm E}^{\{\lambda\}}(\Phi)
:=
(4\pi\lambda)^{-1}\delta(\Phi^2-\lambda)
\ee
can be seen as fundamental. 

However, if, in the presence of a space with Minkowski metric, one
wants to work in a time ordered formalism also in the theory with the
background a different choice for the basis is better adapted. Noticing
that:
\be
2\pi{\mathrm i}\delta(\Phi^2-\lambda)
=
S^-_\lambda(\Phi)-S^+_\lambda(\Phi),
\ee
where
\be
S^\pm_\lambda(\Phi)
=
(\Phi^2-\lambda\pm i\epsilon)^{-1}
\label{scalar}
\ee
with $+$ ($-$) is the time ordered (anti time-orderded) propagator of a
scalar with the squared mass equal to $\lambda$. In the framework of a
time-ordered formalism \mbox{$S^+_\lambda(\Phi)$} could be seen as the 
elementary weight. 

However, with a Minkowski metric---apart from the fact that the case 
\mbox{$\Phi=0$} is not included in the function $w_{\mathrm M}(\Phi^2)$ and 
has to be added separately---the hyperboloid pair characterised by 
\mbox{$\Phi^2=\mathrm{const.}$} has infinite content. Thus with one single 
elementary weight 
the normalisation condition (\ref{normalisation}) cannot be satisfied.
Further, even the difference in content between two hyperboloid pairs is in 
general infinite whereby a superposition of two weights does not suffice to
satisfy the normalisation condition in a non-trivial way. For these reasons 
the minimal construction has to be:
\be
w_{\mathrm M}(\Phi^2)=\sum_{j=1}^{3}a_j S^+_{\lambda_j}(\Phi),
\label{elementaryminkowski}
\ee
with 
\be
\sum_{j=1}^3a_j=0,
\label{condition1}
\ee
and
\be
\sum_{j=1}^3a_j\lambda_j=0.
\label{condition2}
\ee
Then the normalisation condition (\ref{normalisation}) becomes \footnote{If 
the first two conditions are taken into account, the normalisation
condition can be expressed in terms of logarithms of ratios of $\lambda_j$.}:
\be
\frac{4\pi^2}{4}\sum_{j=0}^3a_j\lambda_j\ln{\lambda_j}=1-c.
\label{condition3}
\ee
The conditions (\ref{condition1}) to (\ref{condition3}) can be derived by 
putting a Fourier phase into the normalisation integral (\ref{normalisation}) 
and letting the variable conjugate to $\Phi$ go to zero afterwards. The 
conditions follow from requiring that the limit exist. Then, in general, it 
will also be non-zero [see Eq.~(\ref{condition3})].

As a consequence of these conditions, $w_{\mathrm M}$, as opposed to 
$w_{\mathrm E}$, cannot be positive definite. Condition (\ref{condition1}) 
resembles the one used in Pauli-Villars regularisation.

Any discrete or continuous superposition of delta weights or (time ordered) 
scalar propagators (\ref{scalar}), respectively, fulfilling the normalisation
condition (\ref{normalisation}) is an allowed weight function, 
but in what follows we will concentrate on the minimal forms given in the 
previous equations.

% % % % % % % % % % % % % % % % % % % % % % % % % % % % % % % % % % % % 

\subsection{Generating functional}

The generating functional for the time ordered Green function of QED is
given by:
\be
Z
=
Z_{\mathrm int}
Z_A
Z_\psi,
\ee
with the interaction:
\be
Z_{\mathrm int}
=
\exp\left\{-i\int_x\delta_\eta\hskip -0.7mm\not\hskip -0.5mm\delta_J\hskip 0.5mm\delta_\eta\right\},
\ee
the bosonic:
\be
Z_A
=
\exp\left\{\frac{i}{2}\int_{x,y} J(x)\cdot\Gamma_0(x-y)\cdot J(y)\right\},
\ee
and the fermionic part:
\be
Z_\psi
=
\exp\left\{-i\int_{x,y}\bar{\eta}(x) G_0(x-y)\eta(y) \right\},
\ee
with functional derivatives $\delta$ with respect to the currents 
\mbox{$J,~\eta,~\bar{\eta},$} and the free time-ordered propagators for the 
bosons $\Gamma_0$ and the fermions $G_0$, respectively. 

The modified theory's generating functional ${\cal Z}$ is obtained by 
shifting the gauge field $A$ by $\Phi$ and subsequent averaging with the 
weight $W$. Here this amounts to a modification of the fermionic part leading 
to: 
\be
{\cal Z}
=
Z_{\mathrm int}
Z_A
{\cal Z_\psi},
\ee
where
\be
{\cal Z}_\psi
=
\left<
\exp\left\{
-i\int_{x,y} 
\bar{\eta}(x) G_\Phi(x-y)\eta(y) 
\right\}
\right>_W.
\ee
The other two factors of the generating functional
$Z_{\mathrm int}$ and $Z_A$ can always be taken inside the averaging
integral.
$G_\Phi$ is the time ordered fermion propagator in the field $\Phi$. 
Under the usually made assumption \cite{hp,ls} that all other
condensates are absent it
obeys the equation of motion:
\be
[i\ssh\partial(x)+\SSH\Phi-m]G_\Phi(x-y)=\delta^{(4)}(x-y)
\ee
which is solved by:
\be
G_\Phi(z)=e^{i\Phi\cdot z}G_0(z).
\label{solution}
\ee
Remember that the fermionic propagator in the presence of a medium 
resulting in a chemical 
potential $\mu$ reads \mbox{$e^{i\mu z_0}G_0(z)$}, i.e., technically the 
chemical potential corresponds to the temporal component of a vector and
physically to a conserved charge.
Carrying out the $\Phi$ integral corresponds to a Fourier transformation of
the weight function:
\be
{\cal Z}_\psi
&=&
\sum_{n=0}^{\infty}
\frac{(-i)^n}{n!}
\int_{\{x_m\},\{y_m\}}
\widetilde W(z_n)
\times
\nn
&&\times
\prod_{m=0}^n
[\bar{\eta}(x_m)G_0(x_m-y_m)\eta(y_m)],
\ee
where $\widetilde W(z_n)$ is the Fourier transformation of the weight function
evaluated at \mbox{$z_n:=\sum_{m=0}^n(x_m-y_m)$}. Thence and due to the
Fourier integral the prefactor can be seen as a Lorentz invariant
superposition of chemical potential-like factors.

Eq.~(\ref{solution}) has also similarities with the expressions occuring in
the context of twisted boundary conditions on compact spaces which are used
in lattice calculations \cite{b}. There only the spatial components of the 
vector are non-zero.

The special form of the generating functional leads to the following relation 
for the (higher) correlators:
\be
\left<\langle 0|T\prod_{m=1}^n\psi(x_m)\bar{\psi}(y_m)|0\rangle\right>_W
=
\nn
=
\widetilde W(z_n)\langle 0|T\prod_{m=1}^n\psi(x_m)\bar{\psi}(y_m)|0\rangle
,
\label{factor}
\ee
which remains essentially the same if bosonic operators are added. 
Eq. (\ref{factor}) evidences why $\widetilde W$ has to be time ordered, if the 
functional is to generate time ordered Green functions. 

Even if the second factor on the
right-hand side of the previous equation should show gaussianity---on
a given level---, i.e.,
factorise into two point correlators, the first factor is a
genuine $2n$-point function. Therefore the new theory is not
gaussian. 

Through the limitation to pure gauge configurations of the
background, i.e. such with vanishing field tensor,
the modified theory can be interpreted as one with a non-trivial vacuum
structure without background energy density. One could write:
\be
\left<\langle 0|T\prod_{m=1}^n\psi(x_m)\bar{\psi}(y_m)|0\rangle\right>_W
=:
\nn
=:
\langle\Omega|T\prod_{m=1}^n\psi(x_m)\bar{\psi}(y_m)|\Omega\rangle
,
\ee
where \mbox{$|\Omega\rangle$} stands for the new vacuum. The vacuum
expectation values of the new theory 
\mbox{$\langle\Omega|{\cal O}|\Omega\rangle$} are the averaged vacuum
expectation values \mbox{$\langle\langle 0|{\cal O}|0\rangle\rangle_W$} of 
the old theory.

For two-point functions Eq.~(\ref{factor}) together with the elementary weight
(\ref{elementaryminkowski}) makes the modification of the propagator look like 
a contribution of 
a scalar to the self energy of the fermion without external fermion legs. The
superposition of multiple scalars with different "mass squares" $\lambda$
leads to the summation of the related self-energy bubbles. If, for example,
a three-point function was constructed by including a gauge boson in the
previous correlator, the modification would look like a vertex correction
but still without the outer fermion legs. For
correlators with more than two external fermions the correspondence to
standard scalar loops does no longer persist, whence the new theory is not
identical to one, where the terms for a scalar degree of freedom are added to 
the lagrangian density.

% % % % % % % % % % % % % % % % % % % % % % % % % % % % % % % % % % % % % %

\subsection{The time ordered fermion propagator}

Now we study the Green function central to the modified theory, i.e., the
fermionic two-point function to all orders in $\Phi$.

%  %  %  %  %  %  %  %  %  %  %  %  %  %  %  %  %  %  %  %  %  %  %  %  %  %

\subsubsection{Euclidean metric}

With a euclidean metric the details of the $\epsilon$-prescription are not
important. Therefore the elementary weight of choice is the 
normalised delta weight \mbox{$w_E^{\{\lambda\}}(\Phi^2)$} and $c=0$. As 
mentioned before \mbox{$\Phi^2=0$} in $w_{\mathrm E}$ also means 
\mbox{$\Phi=0$}. However, assuming that $w_{\mathrm E}$ is not divergent at 
this point, this contribution is negligible. Denoting this special averaging 
procedure by \mbox{$\langle{\cal O}\rangle_\lambda^{\mathrm E}$} we get:
\be
\langle G_\Phi(z)\rangle_\lambda^{\mathrm E}
=
\frac{\sin\sqrt{\lambda z^2}}{\sqrt{\lambda z^2}}
G_0(z). 
\label{coordinate}
\ee
This shows that, apart from an oscillatory behaviour, the propagator is
suppressed over large distances $\sqrt{z^2}$. In the limit of short
distances $\sqrt{z^2}$ the free propagator is recovered.
 
Interestingly, Eq.~(\ref{coordinate}) holds not only for the free propagator 
\mbox{$G_0(z)$} but for the full fermionic propagator, i.e., to all orders in
perturbation theory. That is so, because Eq.~(\ref{solution}) is also
satisfied by the full propagator in the presence of the background $\Phi$.
Therefore, the result that the standard propagator is recovered at small
distances and that the background causes a suppression at long distances.

Back to tree-level, in momentum space we get:
\be
\langle G_\Phi(k)\rangle_\lambda^{\mathrm E}
=
\frac{\ssh k+m}{4\sqrt{k^2\lambda}}
\ln\left|
\frac{(\sqrt{k^2}+\sqrt{\lambda})^2-m^2}{(\sqrt{k^2}-\sqrt{\lambda})^2-m^2}
\right|
+
\nn
+
\frac{\ssh k}{4k^2}
\left[
2
-
\frac{k^2+\lambda-m^2}{2\sqrt{k^2\lambda}}
\ln\left|
\frac{(\sqrt{k^2}+\sqrt{\lambda})^2-m^2}{(\sqrt{k^2}-\sqrt{\lambda})^2-m^2}
\right|
\right]
\ee
One can see that the on-shell pole has been removed from the propagator. It
has been replaced by one proportional to \mbox{$1/\sqrt{k^2}$}. Consequently 
the elementary fermions have been removed from the spectrum of asymptotic
states. 

This result is
similar to the one in \cite{hp,ls}, but where also background configurations
with non-vanishing field tensors were admitted in addition to the pure
gauge configurations used here exclusively. 
In coordinate space the weight chosen in \cite{hp} constrained to pure gauge
backgrounds yields :
\be
\langle G_\Phi(z)\rangle_{\mathrm HP}^{\mathrm E}
=
\exp(-z^2\Lambda^2/4)
G_0(z)
\ee
Here as well the free propagator is recovered at small $z^2$ and damped at 
large $z^2$. Even taking the average with this special weight for $2n$-point 
fermion correlators does not lead to a factorisation into two-point 
correlators (gaussianity).

%  %  %  %  %  %  %  %  %  %  %  %  %  %  %  %  %  %  %  %  %  %  %  %  %  %

\subsubsection{Minkowski metric}

In Minkowski space, if one wants to stick to a time ordered treatment the 
adapted weight function has to be chosen for the additional contribution. 
In coordinate space, taking the weight given by Eq.~(\ref{weight}) with 
Eq.~(\ref{elementaryminkowski}) and \mbox{$c=0$} leads to:
\be
\langle G_\Phi(z)\rangle^{\{\lambda_j\}}_{\mathrm M}
=
4\pi^2
\sum_{j=1}^3
a_j
\sqrt{\lambda_j}
\frac
{{\mathrm K}_1(\sqrt{\lambda_j}\sqrt{-z^2+i\epsilon})}
{\sqrt{-z^2+i\epsilon}}
G_0(z)
\nn
\label{propagatorminkowski}
\ee
At \mbox{$z^2=0$} the prefactor of the free propagator \mbox{$G_0(z)$} in the
previous equation goes to 1 due to Eqs.
(\ref{condition1}) to (\ref{condition3}) with \mbox{$c=0$}. Thence, the 
free propagator is recovered in the limit of small $z^2$. If $c$ is
chosen different from zero, once the free propagator is still reproduced taken
together with the explicitely free contribution. The reproduction of
the free propgator at \mbox{$z^2=0$} is an intrinsic consequence of the need
to normalise.

Like in euclidean space the previous relation also holds for the full
propagator, for the same reason as there. 

If \mbox{$\lambda_j>0~\forall~j\in\{1;2;3\}$}, for 
\mbox{$z^2\rightarrow+\infty$} the envelope
of this function decays proportionally to \mbox{$(z^2)^{(-3/4)}$}; for
\mbox{$z^2\rightarrow-\infty$} proportionally to
\mbox{$(-z^2)^{(-3/4)}\exp[-\sqrt{-\min(\{|\lambda_j|\})z^2}]$}. If
\mbox{$\lambda_j<0~\forall~j\in\{1,2,3\}$} the two cases are exchanged. Thus, 
for large absolute values of $z^2$, 
\mbox{$\langle G_\Phi(z)\rangle^{\{\lambda_j\}}_{\mathrm M}$}
is suppressed relative to a free propagator, which shows that the fermions
cannot propagate over arbitrarily large distances.

In momentum space the form of 
\mbox{$\langle G_\Phi(k)\rangle^{\{\lambda_j\}}_{\mathrm M}$} can
be determined best by making use of its correspondence to the one-loop
contribution of a scalar to the self-energy of the fermion. One obtains:
\be
\langle G_\Phi(k)\rangle^{\{\lambda_j\}}_{\mathrm M}
=
\nn
=
i\pi^2
\sum_{j=0}^3a_j
\int_0^1dx
(x\!\ssh k+m)
\ln|(x-x^+_j)(x-x^-_j)|
\ee
with
\be
x_j^\pm
=
\frac{\lambda_j+k^2-m^2}{2k^2}
\pm
\sqrt{\left(\frac{\lambda_j+k^2-m^2}{2k^2}\right)^2+\frac{\lambda_j}{k^2}},
\ee
\mbox{$\forall~j\in\{1;2;3\}$}. For \mbox{$\lambda_j>0~\forall~j\in\{1;2;3\}$} 
the $x$-integration can be carried out yielding:
\be
\int_0^1dx\ln|x-x_j^\pm|
=
\nn
=
(1-x_j^\pm)\ln|1-x_j^\pm|+x_j^\pm\ln|x_j^\pm|-1
\ee
and
\be
\int_0^1x~dx\ln|x-x_j^\pm|
=
\nn
=
\frac{1-(x_j^\pm)^2}{2}\ln|1-x_j^\pm|
+
\frac{(x_j^\pm)^2}{2}\ln|x_j^\pm|
-\frac{x_j^\pm}{2}-\frac{1}{4}.
\ee

\mbox{$\langle G_\Phi(k)\rangle^{\{\lambda_j\}}_{\mathrm M}$} is free of poles.
For small $k^2$ and small mass $m^2$, the propagator becomes:  
\be
\langle G_\Phi(k)\rangle^{\{\lambda_j\}}_{\mathrm M}
\approx
i\pi^2
(\ssh k/2+m)
\sum_{j=1}^3a_j\ln|\lambda_j|.
\ee
For large $k^2$, 
\mbox{$\langle G_\Phi(k)\rangle^{\{\lambda_j\}}_{\mathrm M}$} becomes 
proportional to $k^{-2}$ reproducing the behaviour of the free propagator.

Independent of the details, for \mbox{$c=0$} no freely propagating particles
are described by the propagator in an ensemble of pure gauge backgrounds. 
This also explains its aformentioned correspondence to 
self-energy contributions from scalars {\it without} external fermion legs.

%%%%%%%%%%%%%%%%%%%%%%%%%%%%%%%%%%%%%%%%%%%%%%%%%%%%%%%%%%%%%

\section{Summary}

We have studied gauge field theories with restored Lorentz invariance.
Starting out with a manifestly Lorentz invariant field theory, this symmetry
is broken through the inclusion of a (non-trivial) dependence on a
four-vector $\Phi$. In the explicitely investigated examples said vector
plays the r\^ole of a contribution to the gauge field. The symmetry is 
restored 
by defining correlators as average over a Lorentz invariant ensemble of 
vectors. Apart from the original contribution with $\Phi=0$ the additional
term is characterised by a weight function of the only Lorentz invariant 
$\Phi^2$.
Therefore these theories are connected to mass dimension two vector
condensates. The modifications can also be interpreted as a means to include
the effect of a non-trivial Lorentz invariant vacuum structure into the 
original theory. Some of the structures also appear in stochastic field 
theories. The resulting theories do not show gaussianity.

The presence of the background can bar the asymptotically free 
propagation of the matter fields used to write down the lagrangian density, in 
euclidean and Minkowski space. The bare propagator without the background is 
still reproduced at short distances and high momenta. The dressed propagator 
has no on-shell pole and is suppressed at large distances relative to the 
undressed one.
In non-abelian gauge theories also a constant gauge field can contribute to
the field tensor. Commonly, within this setting, ensembles of backgrounds 
are used which contain pure gauge configurations and field configurations 
leading to a non-zero field tensor. Therefore it was not clear {\it a priori} 
to which contributions the observed effect is connected. We find that when 
constraining the ensemble to pure gauge configurations, the effect is still 
present. That the pure gauge configurations do lead to non-trivial phenomena 
can be understood from the mentioned analogy to a chemical potential. 
Remarkably we have been able to show the suppression of the long-range 
propagation not only based on the fermionic propagator to all orders in the 
background and otherwise at tree level but for the full fermionic propagator, 
i.e., to all loops. 

For correlators involving two fermion fields in general and thus especially 
for the fermion propagator the modifications due to the background resemble
scalar loops bridging the fermion lines. The corresponding diagrams do not
carry external fermion legs although they are direct contributions to the
propagator indicating in this way that they do not involve freely
propagating particles.

Some extensions of the standard model violating Lorentz invariance are based
on the concept of non-commutative field theories. As here Lorentz invariance
has been restored, it would be interesting to study the relationship between 
the present approach and non-commutative field theories not breaking Lorentz 
invariance \cite{nonbreak}.

%%%%%%%%%%%%%%%%%%%%%%%%%%%%%%%%%%%%%%%%%%%%%%%%%%%%%%%%%%%%

\section*{Acknowledgments}

The authors are grateful for inspiring and informative discussions
with Carsten Greiner, Andrew D.~Jackson, Kerstin Paech, 
Olivier P$\grave{\mathrm e}$ne, Dirk Rischke, 
Francesco Sannino, and Kim Splittorff as well as for the hospitality of the
Institute for Theoretical Physics of the Johann Wolfgang Goethe-University
in Frankfurt/Main where part of the work for this project has been carried
out.
Thanks are again due to Kim Splittorff for the careful reading of and
useful comments on the manuscript.

%%%%%%%%%%%%%%%%%%%%%%%%%%%%%%%%%%%%%%%%%%%%%%%%%%%%%%%%%%%%

\end{document}